\documentclass{article}

\usepackage[english]{babel}
\usepackage{amssymb}
\usepackage{biblatex}
\usepackage[letterpaper,top=2cm,bottom=2cm,left=3cm,right=3cm,marginparwidth=1.75cm]{geometry}

\usepackage{amsmath}
\usepackage{graphicx}
\usepackage[colorlinks=true, allcolors=blue]{hyperref}
\usepackage{stmaryrd}
\usepackage{algpseudocode}
\usepackage{algorithm}
\usepackage{tikz-cd}
\usepackage{float}
\usepackage{xfrac}

\title{A Numerical System For Nested Spaces - Defining An Intuitive, Universal Coordinate System For Self-Similarity}
\author{Petal Belle Mokryn}

\begin{document}
\maketitle

\begin{abstract}
In this paper I introduce a simple, universal numerical system for describing all self-similar nested spaces. For a nested space made of $N>1$ smaller copies of itself, the location of all points can be specified using the double shift space $\lbrace 0,1,...,N-1 \rbrace^{\mathbb{Z}}$. Connectivity rules are established to fully define the topology of the nested space, and immediate consequences are investigated using the numerical system defined. I introduce the numerical system and demonstrate how it unifies previous existing works under a single, simple framework, to argue for its adoption as standard in discussions of nested fractals.
\end{abstract}

\section{Introduction}
Ever since the 70's, there has been a prevalent interest in self-similarity and fractal spaces, with many famous examples such as the Cantor set and the Sierpinski triangle, Julia sets, and more.\\
Mathematicians and physicists alike have been drawn to fractal mathematics like moths to a flame, enticed by the plethora of new concepts hanging like fruit ripe for the picking, that may be key to understanding many systems.\\
There are applications of fractal mathematics in Lattice Quantum Gravity \cite{ambjorn2011lattice}, condensed matter physics \cite{LIU1986207}, stochastics \cite{bandt1995fractal}, economics \cite{takayasu2000fractal}, and more.\\\\
One of the simplest kinds of fractals, and thus the most studied, are the \textit{nested fractals}. Roughly speaking, that means fractals made of some finite $N>1$ number of smaller copies of themselves. Examples include the Cantor Set, Sierpinski Gasket, Sierpinski Carpet, the Snowflake fractal, and more.\\\\
In this paper, I define a rigorous yet highly general notion of "$N$-parts nested spaces", capturing much of the existing literature on the topic under a single, general framework. I demonstrate that all nested spaces can be simply and effectively described using the double shift space $\lbrace 0,1,...,N-1 \rbrace^{\mathbb{Z}}$ equipped with a suitable topology. I define an equivalence relation I call "connectivity rules", which defines a sort of arithmetic on $\lbrace 0,1,...,N-1 \rbrace^{\mathbb{Z}}$, and fully details the topological structure of nested spaces.\\\\
The numerical system introduced in this paper is a simple, effective way to describe nested spaces, which generalizes all previous notations and constructions, and gives an easy and effective form of notation, reminiscent of positive real numbers.

\section{Background Formulation}

\subsection{Defining Nested Spaces} \label{NestedDef}
Nesting, in this paper, is a topological property. First I will define which sets \textit{can} be nested sets/spaces. Using those definitions, I will then define the \textbf{$N$-parts nested cell-structure topology}. Finally, nested spaces and nested sets will be defined as sets that can be nested, endowed with either the $N$-parts nested cell-structure topology or with the appropriate compact subspace topology of it, respectively.\\
\subsubsection{Definition 1: Nested Sets.}
Let $S$ be a non-empty set. We say $S$ can be a \textit{$N$-parts nested set}, if there exist $N>1$ distinct, non-trivial, non-identity functions $F_{0},...,F_{N-1} : S\to S$, such that each $F_{i}$ is defined on all of $S$, and the following equation holds:
\begin{equation} \label{NestedSet}
    S = \underset{i=0,...,N-1}{\bigcup} F_{i}\left [ S \right ]
\end{equation}
The set $S$, if endowed with and compact under the \textit{nested cell-structure topology} defined later in this paper, will then be called a $N$-parts nested set, and the functions $F_{i}$ will be called its nesting relations. Note that very few restrictions are placed on what kind of functions the nesting relations actually are, as this paper aims to be as general as possible.\\
The following equation holds trivially:
\begin{equation} \label{NestingRelation}
    \forall i=0,...,N-1: F_{i}\left[ S \right] = F_{i}\left [ \underset{j=0,...,N-1}{\bigcup} F_{j}S \right ] = \underset{j=0,...,N-1}{\bigcup} \left ( F_{i}\circ F_{j}\right ) \left [ S \right ]
\end{equation}

\subsubsection{Definition 2: Nested Spaces.}
We say a set $X$ can be a \textit{$N$-parts nested space} if the following conditions are satisfied:\\
\begin{itemize}
    \item $X$ is a countable nested union $X = \underset{n\in\mathcal{A}}{\bigcup} X_{n}$ , where $\mathcal{A}\subseteq \mathbb{N}$, and $k>0\in\mathcal{A}\Rightarrow k-1\in\mathcal{A}$ . 
    \item Each $X_n$ can be a $N$-parts nested set, each with its own nesting relations $\left\lbrace F_{n;i} \right\rbrace_{i=0}^{N-1}$
    \item The following relation holds: $\forall n,n+1\in \mathcal{A}: X_n = F_{n+1 ; 0}\left[ X_{n+1} \right]$
    \item For all $n,n+1\in\mathcal{A} , i \in \lbrace 0,...,N-1 \rbrace: $, the following diagram commutes:
    \begin{equation}
        \begin{tikzcd}
            &X_{n+1} \arrow[d, "F_{n+1;0}"'] \arrow[r, "F_{n+1;i}"] &F_{n+1;i}[X_{n+1}] \arrow[d, "F_{n+1;0}"]\\
            &X_{n} \arrow[r, "F_{n;i}"'] &X_{n}
        \end{tikzcd}
    \end{equation}
\end{itemize}
If $\left| \mathbb{A} \right|<\infty$, then a nested space is also a nested set. It's trivial to show that every nested set is a nested space, but not every nested space is a nested set.

\subsubsection{Defining Connectivity Rules} \label{ConnectivityDef}
Plugging eq. \ref{NestingRelation} into eq. \ref{NestedSet} and iterating, we get the following relation:
\begin{equation} \label{UnionOfPointsDef}
    S = \lim\limits_{M\to\infty}\underset{i_{0},...,i_{M-1}=0,...,N-1}{\bigcup} F_{i_{0}...i_{M-1}}\left [ S \right ]
\end{equation}
Where $F_{i_{0}...i_{M-1}} \left [ S \right ] = \left( F_{i_{0}}\circ...\circ F_{i_{M-1}} \right)\left [ S \right ]$ are a generalization to all possible nesting relations of the "$M$-cells" indexed using "words of length $M$", introduced in (Strichartz 2006)\cite{strichartz2006differential}. In this paper I call them "base-$N$ words of length $M$", to be specific. By virtue of the nesting relations, and the fact that set limits of the form $\lim\limits_{k\to\infty}A_{k} , \forall k:A_{k+1}\subset A_{k}$ always converge, there exist objects of the form
\begin{equation}\label{TopPointsDef}
F_{\omega}=\lim\limits_{M\to\infty}F_{i_{1}...i_{M}}\left [ S \right ]
\end{equation}
And they serve, as I will prove later, the topological role of "points" under the nested cell-structure topology.\\
Now that we have this concept of "points", we can thus define the "connectivity rules" of $S$ as relations of the form
\begin{equation} \label{ConnectivityFormalDef}
    F_{\omega_{1}}=F_{\omega_{2}} \, , \, \omega_1 \neq \omega_2
\end{equation}
Where $F_{\omega_{1}},F_{\omega_{2}}$ are the topological points introduced in eq. \ref{TopPointsDef}. We will be proving theorems about the connectivity rules and their consequences later in this paper, after introducing a much more convenient notation system to discuss nested spaces with.\\\\

\subsubsection{The $N$-Parts Nested Cell-Structure Topology}
Given a set $X$ that can be a nested space, we define its $N$-Parts Nested Cell-Structure Topology $\tau_{\text{Nested}}$ as follows:
\begin{itemize}
    \item All $X_{n}$'s are members of $\tau_{\text{Nested}}$ : $\forall n\in\mathcal{A}: X_{n}\in\tau_{\text{Nested}}$
    
    \item For all $n\in\mathcal{A}$, and all $M\in\mathbb{N}/\lbrace0\rbrace$, all $M$-cells of $X_{n}$ are members of $\tau_{\text{Nested}}$

    \item All arbitrary unions of the sets mentioned above are members of $\tau_{\text{Nested}}$

    \item All finite intersections of the sets mentioned above are members of $\tau_{\text{Nested}}$
\end{itemize}
If $|\mathcal{A}|<\infty$ , then $X$ is compact under $\tau_{\text{Nested}}$, and we call $X$ both a $N$-parts nested space, and a $N$-parts nested set. If $\mathcal{A}=\mathbb{N}$, then $X$ isn't compact under $\tau_{\text{Nested}}$, and we only call $X$ a $N$-parts nested space. All nested sets are nested spaces, but not all nested spaces are nested sets.\\
Note that all objects $F_{n;\omega}$, with $\omega$ a base-$N$ word of length $\infty$, are topologically points under $\tau_{\text{Nested}}$ because they have Lebesgue covering dimension zero.

\subsection{Defining The Numerical System $[T_{N}]_{X}$}
\subsubsection{Defining the $N$-Parts Numbers $T_{N}$} \label{N-PartsNumbers}
Let us define the double shift space $\left\lbrace 0,1,...,N-1 \right\rbrace^{\mathbb{Z}}$, and for all $p\in\left\lbrace 0,1,...,N-1 \right\rbrace^{\mathbb{Z}}$ we use notation $p=...p_{1}p_{0}.p_{-1}p_{-2}...$ , as if we're writing a base-$N$ positive real number. Let us also restrict to $T\subset\left\lbrace 0,1,...,N-1 \right\rbrace^{\mathbb{Z}}$, such that for all $p\in T$, there must exist some $n\in\mathbb{N}$ such that $\forall j > n : p_{j}=0$, meaning that if we were to endow $T$ with the real number line topology, all $p\in T$ would be finite.\\\\
We define the $N$-parts nested cell-structure topology $\tau_{N}$ on $T$ as follows:
\begin{itemize}
    \item For all $k\in\mathbb{Z}$, and all base-$N$ natural numbers $A$, the base-$N$ real number interval $I = N^{k}\cdot \left[ A.\overline{0} , A.\overline{N-1} \right)$ is a member of $\tau_{N}$

    \item All arbitrary unions of such intervals are members of $\tau_{N}$

    \item All finite intersections of such intervals are members of $\tau_{N}$\\
\end{itemize}
The restricted double shift space $T$ endowed with the $N$-parts nested cell-structure topology $\tau_{N}$ is now called the $N$-parts numbers, and marked $T_{N}$.

\subsubsection{Charting $N$-Parts Nested Spaces with $(n,\omega)$}
Let us consider a $N$-parts nested space $X$ with $\mathcal{A}=\mathbb{N}$,\\
and the series of $N$-parts nested sets it's made of $\left\lbrace X_n \right\rbrace_{n=0}^{\infty}$ .\\\\
Let us define a set $\mathbb{K}_{N}$ of all ordered pairs $(n,\omega)$,\\
where $n\in\mathbb{N}$ and $\omega$ is a base-$N$ word of length $\infty$.\\
For all base-$N$ words $\omega = \omega_{0}...\omega_{M-1}$ of finite length $M<\infty$ , we can amend $\omega$ to be a base-$N$ word of length $\infty$ by changing it to $\omega = \omega_{0}...\omega_{M-1}0000....$\\
The ordered pairs $(n,\omega)\in\mathbb{K}_{N}$ now map all points in $X$ using the surjective mapping
\begin{equation}
    \tilde{F}_{X}(n,\omega) = F_{n;\omega}[X_n]
\end{equation}

\subsubsection{Charting $N$-Parts Nested Spaces using $T_{N}$}
We can now define another surjective mapping, $G(n,\omega): \mathbb{K}_{N} \to T_{N}$, defined as follows:\\
Given $(n,\omega)\in \mathbb{K}_{N}$, we write out explicitly $\omega = \omega_{0}\omega_{1}...$.
Let us define $G(n,\omega) = p$, such that $p = ...p_{1}p_{0}.p_{-1}p_{-2}...\in T_{N}$, and for each specific digit $\omega_{k},k\in\mathbb{N}$ of $\omega$, the digits of $p$ are assigned as follows:\\
\begin{itemize}
    \item $\forall k\in\mathbb{N}: p_{n-k}=\omega_{k}$

    \item $\forall j>n: p_{j}=0$
\end{itemize}
The mapping $G$ is not invertible, because there are different $\mathbb{K}_{N}$, $(n_1 , \omega_1)\neq (n_2 , \omega_2)$ that get mapped to the same $p\in T_{N}$. A limited inverse $\tilde{G}^{-1} : T_{N} \to \mathbb{K}_{N}$ can still be defined:\\
Given any $p\in T_{N}$, let $n$ be the largest natural number such that the digit $p_{n}$ of $p$ is non-zero. If such a number doesn't exist, then let $n=0$. We now define $\tilde{G}^{-1}(p) = (n,\omega)$ where the digits of $\omega$ are assigned $\forall k \in\mathbb{N}:\omega_{k} = p_{n-k}$\\\\
The "inverse" $\tilde{G}^{-1}$ now allows us to define a surjective mapping
\begin{equation}
    M_{X}=\left(\tilde{F}_{X} \circ \tilde{G}^{-1}\right): T_{N} \to X
\end{equation}
From $T_{N}$ to any $N$-parts nested space $X$, allowing us to chart $X$ with coordinates $T_{N}$.

\subsubsection{The Quotient Space $[T_{N}]_{X}$ Homeomorphic to $X$}
The coordinate chart $M_{X}:T_{N}\to X$ , on its own, doesn't give us a complete topological description of $X$. To do that, we also need to describe the connectivity rules of $X$ in $T_{N}$.\\
Let us define on $T_{N}$ an equivalence relation $\stackrel{X}{\sim}$ , such that
\begin{equation}
    \forall a,b\in T_{N}: a \stackrel{x}{\sim} b \Leftrightarrow M_{X}(a)=M_{X}(b)
\end{equation}
For any $N$-parts nested space $X$, we call the equivalence relation $\stackrel{x}{\sim}$ defined on $T_{N}$ by $X$ the \textit{Connectivity Rules} of $X$.\\\\
Let us now define the quotient space $[T_{N}]_{X} \triangleq \sfrac{T_{N}}{\stackrel{x}{\sim}}$ , equipped with the quotient space topology $\sfrac{\tau_{N}}{\stackrel{x}{\sim}}$ .\\
Restricting $M_{X}$ to $[T_{N}]_{X}$, the mapping $M_{X}:[T_{N}]_{X} \to X$ is a bijection, and note that the nested cell-structure topology $\tau_{\text{Nested}}$ is exactly the final topology of $\sfrac{\tau_{N}}{\stackrel{x}{\sim}}$ induced on $X$ by $M_{X}$, making $M_{X}$ a homeomorphism.\\\\
In conclusion, for all $N$-parts nested spaces with $\mathcal{A}=\mathbb{N}$:
\begin{equation}
    X \cong [T_{N}]_{X}
\end{equation}
Under homeomorphism $M_{X}:[T_{N}]_{X} \to X$\\\\
To chart a nested set with $|\mathcal{A}| = K <\infty$,\\
Simply restrict $T_{N}$ to the shift space $\lbrace 0,1,...,N-1 \rbrace^{\lbrace K, K-1, ... \rbrace} \subset T_{N}$ before taking the quotient by $\stackrel{x}{\sim}$

\subsection{Arithmetic Operations on $T_{N}$ and Their Implications on $[T_{N}]_{X}$}
Lastly, we need to consider arithmetic operations on $T_{N}$ , and their implications on $[T_{N}]_{X}$. The usual arithmetic operations on base-$N$ positive real numbers \textit{can} be used, but they may not have the same interpretation for general nested spaces as they have for the positive real numbers in ordinary use. In fact, they can be completely nonsensical for some nested spaces.\\
There are two operations that \textit{do} have a simple, intuitive interpretation for all nested spaces.
\subsubsection{Scaling}
The first operation we define is multiplication by integer powers of $N$, which we will call according to its intuitive interpretation, \textbf{scaling}, or \textbf{shift}. 
\begin{equation}
    \text{Sc}:T_{N} \times \mathbb{Z} \to T_{N}:\; \text{Sc}(a,k)=N^{k}\cdot a
\end{equation}
We'll simply mark the scaling operation using "$N^{k}\cdot$" or "$\cdot N^{k}$" in future use, for simplicity. Note that the scaling operation $\cdot N^{k}$ for all $k\in\mathbb{Z}$ preserves connectivity rules, meaning $a\stackrel{x}{\sim} b \Leftrightarrow N^{k}\cdot a \stackrel{x}{\sim} N^{k}\cdot b$, and that the scaling operation also defines an equivalence relation on $[T_{N}]_{X}$.\\

\subsubsection{Imposition}
Another operation with clear intuitive interpretation, is one we'll have to define algorithmically. I call this operation \textit{imposition}. $a \leftslice b$, or \textit{$a$ impose $b$}, and it's a set-valued operation defined by the following pseudocode:
\begin{algorithm}[H]
\begin{algorithmic}
    \Function{$\leftslice$}{$a,b$}
        \State define $tempA$ := $a$
        \State define $tempB$ := $b$
        \State define $ConnectedA$ := $\left\lbrace  p \in T_{N} \Big| a \stackrel{x}{\sim} p \right\rbrace$
        \State define $ConnectedB$ := $\left\lbrace  p \in T_{N} \Big| b \stackrel{x}{\sim} p \right\rbrace$
        \State define $Imposed = \lbrace  \rbrace$
        \ForAll{$A \in ConnectedA$}
            \ForAll{$B \in ConnectedB$}
                \State define $temp$ := $A$
                \ForAll{$i\in\mathbb{Z}$}
                    \If{$A$.digits[$i$] == 0}
                        \State $temp$.digits[$i$] := B.digits[$i$]
                    \EndIf
                \EndFor
                \State $Imposed$.AddMember[$temp$]
            \EndFor
        \EndFor
        \State \Return $Imposed$
    \EndFunction
\end{algorithmic}
\end{algorithm}
The intuitive meaning of $a\leftslice b$ is "start at $a$, and 'walk in the direction' of $b$ in all ways the connectivity rules of the nested space $X$ allow".\\
Note that the operation of imposition also preserves connectivity rules, meaning that $\forall a,b,c \in T_{N}: a \stackrel{x}{\sim} b \Rightarrow (a\leftslice c) = (b\leftslice c)$\\
In actual usage, we likely won't be working with the entire set $a\leftslice b$, but rather only with specific useful members of it.

\section{Consequences}
\subsection{Properties of Nested Spaces and Connectivity Rules}
\begin{enumerate}
    \item Given a connectivity rule $a \stackrel{x}{\sim} b$ , where $a , b \in T_{N}$, it's easy to prove $\forall k\in\mathbb{Z}: (N^k \cdot a) \stackrel{x}{\sim} (N^k \cdot b)$ as they are essentially the same connectivity rule, either in different $X_n$'s or in different M-cells of $X_0$.\\
    Such connectivity rules are to be recognized as equivalent to each other, so we define the set of connectivity rules of a $N$-parts nested space modulo this equivalence.
    We define the set of connectivity rules of a nested space $X$ as the quotient set:
    \begin{equation}
        C_{X} = \sfrac{\left(\sfrac{\left\lbrace \lbrace a,b \rbrace | a,b\in T_{N} , a \stackrel{x}{\sim} b \right\rbrace}{\left\lbrace \lbrace p,p \rbrace | p\in T_{N}  \right\rbrace}\right)}{\cdot N^{k}}
    \end{equation}
    And we call its cardinality $|C_{X}|$ the number of connectivity rules in $X$. Note that if $|C_{X}|<\infty$, then $X$ is post-critically finite.

    \item Given a $N$-parts nested space $X$, a \textit{set of connected points} in $X$ is any set $S$ of unordered pairs $\lbrace a,b \rbrace$, such that $a,b\in T_{N}$ and $a \stackrel{x}{\sim} b$ .\\\\
    We call a \textit{Basis of Connectivity Rules} of $X$, any set $G$ of connected points in $X$ such that all of $C_{X}$ can be defined from $G$ using the operations of scaling and imposition.\\\\
    Finally, we call a \textit{Generator of Connectivity Rules} $G_{X}$ of $X$, any basis of connectivity rules $G_{X}$ of $X$, such that for all other basis $G'$ of connectivity rules of $X$, $|G_{X}|\leq |G'|$ .

    \item \textbf{Hypothesis:}\\
    For all $N>1$, there exists some finite $K(N)\in\mathbb{N}$ , such that for all $N$-parts nested spaces $X$ :
    \begin{equation}
        |G_{X}|\leq K(N)
    \end{equation}
    Should this hypothesis prove to be true, I will call a $N$-parts nested space $X$ with $|G_{X}| = K(N)$ \textit{Maximally Connected}.

    \item Given a $N$-parts nested space $X$ with $\mathcal{A}=\mathbb{Z}$, for all $i\in \lbrace 0,1,...,N-1 \rbrace , k\in\mathbb{Z}$, we call the interval $I_{i,k} = N^{k}\cdot \left[ i.\overline{0} , i.\overline{N-1} \right)$ the \textit{$i$-th part} of $X$ at scale $k$.\\\\
    We call $I_{i,k}$ an \textit{edge part} of $X$ if the connectivity rule $0.\overline{i} \stackrel{x}{\sim} i.\overline{0}$ holds.\\\\
    If no such connectivity rule holds, but other types of connectivity rules \textit{do} exist for $I_{i,k}$, then I call $I_{i,k}$ a \textit{middling part} of $X$ at scale $k$.

    \item \textbf{Hypothesis:}\\
    If a $N$-parts nested space $X$ is path-connected, then every middling part $I_{a,k}$ of $X$ at scale $k$ has at least two connectivity rules, such that $I_{a,k}$ is connected to at least two other parts of $X$ at scale $k$.

    \item \textbf{Ease of usage:}\\
    $T_{N}$ charts cells of any size, anywhere in the $N$-parts nested space, using real number intervals of the form $I = N^{k}\cdot \left[ A.\overline{0} , A.\overline{N-1} \right)$
    
    \item \textbf{Ease of usage:}\\
    For a nested space $X$, existing literature often defines vertex sets $V_{...}$ as the boundaries of a corresponding $M$-cell $F_{...}$ of a specific $X_{n}$ . In $[T_{N}]_{X}$, that simply means taking the boundary under topology $\sfrac{\tau_{N}}{\stackrel{x}{\sim}}$ of the real number intervals $I = N^{k}\cdot \left[ A.\overline{0} , A.\overline{N-1} \right)$
\end{enumerate}

\subsection{Example Nested Space: The Positive Real Number Line} \label{ExampleR}
The positive real number line can be defined in a way similar to how (Strichartz 2006)\cite{strichartz2006differential} defined the unit interval, but more straightforward:\\
We define the positive real number line as a 2-parts nested space charted by $T_{2}$, with the connectivity rule $0.\overline{1} \sim 1.\overline{0}$. The arithmetic operation of addition then trivially follows from this connectivity rule used together with scaling and imposition.\\\\
In fact, the positive real number line and its arithmetic can also be defined as a $N$-parts nested space of \textit{any} natural $N>1$, using $T_{N}$ together with the connectivity rule $\forall i = 0,...,N-2: i.\overline{(N-1)} \sim (i + 1).\overline{0}$.\\\\
As a matter of fact, for $N=10$ we get\\
an incredibly familiar-looking connectivity rule, $0.\overline{9} \sim 1$ !\\\\
Furthermore, by restricting the number of digits either to the left or to the right of the decimal dot on $[T_{N}]_{\mathbb{R}_{+}}$, one can easily define either finite intervals on $\mathbb{R}_{+}$ or the natural numbers, respectively.

\subsection{Example Nested Space: The Extended Sierpinski Triangle}\label{ExampleSG}
The infinitely extended Sierpinski triangle with infinite detail, like those shown by Fukushima \cite{fukushima1992spectral} and Shima \cite{shima1992lifschitz}, can be defined as a 3-parts nested space charted by $T_{3}$ with the connectivity rules $\forall i,j = 0,1,2: i.\overline{j} \sim j.\overline{i}$\\\\
There are two other popular constructions in literature of the Sierpinski triangle:\\
\begin{enumerate}
    \item A construction like that done by Strichartz \cite{strichartz2006differential}, analogous to the unit interval on $\mathbb{R}_{+}$ ; a 3-parts nested set Sierpinski triangle, compact in $\mathbb{R}^2$ and containing infinite detail.

    \item A construction like that done by Rammal \cite{rammal1982spectrum}, made by joining together triangles of defined elementary size to form a finitely detailed but infinitely extended Sierpinski triangle in $\mathbb{R}^2$, analogous to the natural numbers.
\end{enumerate}
Both of these constructions are easily achieved from our definition of the 3-parts nested space Sierpinski triangle above, simply by restricting the number of digits to the left or to the right of the decimal point, respectively, in exactly the same way we defined the finite interval and natural numbers in Sec. \ref{ExampleR}.

\section{What's Next?}
There is much work left to be done on this project of mine.
\subsection{Scope: Undergrad Project}
First, should I find an advisor this semester, I would like to start by engaging with my hypothesis, and by simply translating the existing body of work on the dynamics of nested fractals to my $T_{N}$ notation.\\
That means, doing either part or all of the following list:
\begin{itemize}
    \item Trying to prove my hypothesis shown above, and any other useful insights.
    
    \item Translating and generalizing the laplacian on nested fractals defined by the likes of Kigami \cite{kigami2001analysis}, Strichartz \cite{strichartz2006differential}, Shima \cite{shima1992lifschitz} and more to my coordinate system $T_{N}$

    \item Translating the method of \textit{spectral decimation}, shown by the likes of Rammal \cite{rammal1982spectrum}, Strichartz \cite{strichartz2006differential} and more as the go-to method in existing literature to solving differential equations on fractals, to my coordinate system $T_{N}$

    \item Translating the eigenfunctions of the laplacian on nested fractals, the harmonic functions in particular, to their expressions as functions of $[T_{N}]_{X}$

    \item Translating and generalizing the fractal Green's functions shown by Kigami \cite{kigami2000green} to their expressions as functions of $T_{N}$

    \item Engaging with work on the properties of dynamics on fractals, such as that done by Akkermans \cite{akkermans2012spatial}.

    \item Translating the gradient and harmonic/"finite energy" coordinates shown by the likes of Tepylaev \cite{teplyaev2008harmonic}, Kelleher \cite{kelleher2017differential} and Hinz \cite{hinz2015finite}.

    \item Comparing and contrasting the cases of analysis on finitely and infinitely ramified (see Barlow \cite{barlow1999brownian}) fractals using the language of $T_{N}$, looking for any useful insights that may arise. 
\end{itemize}

\subsection{Scope: Advanced Degree}
In further future work, potentially as a Masters/PhD project, I would also like to try the following:
\begin{itemize}
    \item Engage with and try to generalize the work on differential forms on fractals, done by Kelleher \cite{kelleher2017differential}, Aaron \cite{aaron2012hodge}, Cipriani \cite{cipriani2011differential}, and Hinz \cite{hinz2012vector} from the perspective of $T_{N}$ notation.

    \item Use $T_{N}$ construction and notation to try and generalize the field of algebraic topology as a whole, cohomology in particular, to work on nested spaces.\\
    I wish to do so by defining an equivalent to simplexes and chain-complexes, either on $T_{N}$, or on some specific $[T_{N}]_{X}$'s with unique properties, possibly the \textit{maximally connected} $N$-parts nested spaces, if they do indeed exist.

    \item Try and engage with further work on the topic from the perspective of $T_{N}$ coordinates, such as that done by Hu \cite{hu2001nonlinear}, David \cite{david2017note}, Ionescu \cite{ionescu2012derivations}, Berth{\'e} \cite{berthe2011fractal}, Bedford \cite{bedford1986dimension}, and more.
\end{itemize}

\section*{Acknowledgements:}
I would like to thank Prof. Eric Akkermans for introducing me to the topic of analysis on nested fractals. You've enriched me and helped me study this topic, as well as mathematics in general, to a level beyond what I could've otherwise learned in undergrad.

\printbibliography
\end{document}